\documentclass[%
preprint,
 amsmath,amssymb,
 aps,
prb,
sort&compress,numbers
]{revtex4-2}

\usepackage{graphicx}
\usepackage{dcolumn}
\usepackage{bm}
\usepackage{comment}
\usepackage{makecell}

\begin{document}

\preprint{APS/123-QED}

\title{Computationally efficient method for calculating electron-phonon coupling for high-throughput superconductivity search}

\author{Oliver A. Dicks}
\email[Corresponding author. Email address: ]{oliver.dicks@ubc.ca}
\affiliation{Stewart Blusson Quantum Matter Institute, University of British Columbia}

\author{Kateryna Foyevtsova}
\affiliation{Stewart Blusson Quantum Matter Institute, University of British Columbia}

\author{Ilya Elfimov}
\affiliation{Stewart Blusson Quantum Matter Institute, University of British Columbia}

\author{Rohit Prasankumar}
\affiliation{Deep Science Fund, Intellectual Ventures, Bellevue, Washington,United States}

\author{George Sawatzky}
\affiliation{Stewart Blusson Quantum Matter Institute, University of British Columbia}

\begin{abstract}
Using a computationally inexpensive frozen phonon approach we have developed a technique which can be used to screen large unit cell materials and systems for enhanced superconducting critical temperatures. The method requires only density functional theory (DFT) calculated electronic band structures of phonon modes corresponding to atomic displacements for various materials. We have applied this method to well known conventional superconductors including MgB$_2$, H$_{3}$S and other hydrides as examples.

\end{abstract}

\maketitle

\section{\label{sec:Intro} Introduction}

High throughput structure searches for superconductors with critical temperatures ($T_{c}$) approaching 300K continue apace\cite{Shipley2021,Zhang2020}, with huge amounts of time and computational resources being dedicated to this task. Concurrently, recent developments in machine learning techniques \cite{Merchant2023} have unlocked a new order of magnitude in the number of novel stable materials that can be screened for properties, including superconductivity. However, calculations of $T_{c}$ or other material properties associated with superconductivity, like the electron-phonon coupling constant (EPC), $\lambda$, are expensive to compute using accurate first order perturbation theory such as Migdal-Eliashberg, implemented in codes like EPW \cite{Noffsinger2010}. These calculations become entirely unfeasible when screening potential superconducting systems with large numbers of atoms, such as large primitive cell crystal structures, doped systems, many layered structures or 2-dimensional thin films on substrates.

Using a computationally inexpensive frozen phonon approach we have developed a technique that can approximate the calculation of $\lambda$ for individual phonon modes from DFT (Density Functional Theory) single electron band structures that improves on previous frozen phonon methods \cite{Sun2022,Yin2013}. This can be applied to the task of filtering potential high-$T_{c}$ superconducting materials from large data sets. Sun et al. \cite{Sun2022} calculate the contribution to the EPC from only the zone-center ($\Gamma$-point) modes and assume that they are representative of the EPC over the whole Brillouin zone. This method does demonstrate a relationship between the EPC of $\Gamma$-point modes and that calculated over the whole Brillouin zone (BZ), but a phase space or``fudge-factor" (to quote the authors) of 0.22 has to be applied to the estimated EPC. In Yin et al.'s \cite{Yin2013} work they instead use frozen-phonons to go beyond the standard LDA (local density approximation) and GGA (general gradient approximation) functionals to calculate the changes of the electronic structure for specific phonons of interest in superconducting materials using more accurate, but expensive, hybrid functionals and GW methods. 

In this paper we outline a method using the effect of ionic displacements on the electronic band energies near the Fermi energy to predict the strength of electron-phonon coupling. We show that, not only is this method able to predict which specific phonon modes contribute strongly to the EPC in a given material, but that it is also able to predict $\lambda$ with sufficient accuracy to be used to pre-screen materials. It allows the calculation of the contribution to $\lambda$ of specific phonon modes of systems with large numbers of atoms, including 2D surfaces and interfaces, where full Eliashberg calculations are infeasible.

\section{\label{sec:Methodology} Methodology}

We used PHONOPY \cite{phonopy-phono3py-JPCM,phonopy-phono3py-JPSJ} to calculate phonon displacement vectors with the DFT code Quantum ESPRESSO \cite{Giannozzi2017} as the calculator, and generate the corresponding frozen phonon cells. Quantum ESPRESSO \cite{Giannozzi2017} was then used to calculate the electronic band structures of these frozen phonons in a variety of materials iso as to estimate their electron-phonon coupling strength, $\lambda$.

The GGA Perdew-Burke-Ernzerhof (PBE) functional was used for all systems, with k-mesh densities of approximately 0.001 \AA$^3$ (e.g. a 24$\times$24$\times$24 k-mesh for a primitive cell of H$_3$S and a 12$\times$12$\times$12 k-mesh for a 2$\times$2$\times$2 supercell). The PAW (projector augmented-wave) pseudopotentials \cite{kucukbenli2014projector} were used for all calculations in order to allow comparisons between our systems. All input files and code are available for download \cite{git}.

\subsection{\label{sec:gandlamdamethod}Approximating the electron phonon coupling}

In order to perform a high throughput materials’ search for high temperature superconductors we have to be able to calculate the superconducting critical temperature. A common approach is to use the Allen-Dynes formula to calculate T$_{c}$ 
\begin{equation} \label{eq:Allen-Dynes}
T_c=\frac{\omega_{log}}{1.20}\exp{\left(-\frac{1.04\left(1+\lambda\right)}{\lambda-\mu^\ast\left(1+0.62\lambda\right)}\right)} .
\end{equation}
where the main problem is the determination of the dimensionless electron-phonon coupling $\lambda$, since the prefactor ($\omega_{log}$ is the logarithmic average of the phonon frequency) and $\mu^\ast$ (the Coulomb pseudopotential taking into account electron repulsion) can subsequently be easily calculated or approximated.

We therefore introduce a ``poor man's" methodology to calculate an approximation of the Migdal-Eliashberg expression for the electron-phonon coupling (as calculated by, for example, EPW\cite{Noffsinger2010} in Quantum ESPRESSO), which appears in the Allen-Dynes equation. Similar approaches have also used frozen phonons to estimate $\lambda$\cite{Yin2013,Sun2022}, but the method described here is an approximation which allows similar speedup whilst also including the effects of modes outside the $\Gamma$ point.

A general expression for the electron phonon coupling (EPC) strength omitting band indices, $\lambda$, can be written as \cite{Meregalli1998};
\begin{equation} \label{eq:lamda}
    \lambda = \frac{2}{N(\varepsilon_{F})N_{q}}\sum_{k,q,\nu} \lvert M_{k,k+q}^{\nu} \rvert^{2} \delta(\varepsilon_{k}-\varepsilon_{F})\delta(\varepsilon_{k+q}-\varepsilon_{F})
\end{equation}
where $N(\varepsilon_{F})$ is the density of states at the Fermi level, $\varepsilon_{F}$, $N_{q}$ is the number of $q$ points sampled and $\varepsilon_{k}$ is the energy eigenvalue of the band at wave vector $k$. The full expression for the electron-phonon matrix elements is given by
\begin{equation} \label{eq:M}
    M_{k,k+q}^{\nu} = \sum_{j} \left( \frac{1}{\sqrt{2M_{j}\omega_{q, \nu}}} \right) \langle k+q|\delta V / \delta u_{q,j}^{\nu}|k \rangle
\end{equation}
where $j$ indexes the sum over all atoms in the cell, $M_j$ is the mass of atom $j$, $\omega_{q, \nu}$ is the phonon frequency of phonon branch $\nu$ and wave vector $q$, $u_{q,j}^{\nu}$ is the displacement vector of atom $j$ for a given phonon and $\delta V / \delta u_{q,j}^{\nu}$ is the partial derivative of the total Kohn-Sham potential energy with respect to that displacement.

\begin{figure*}
\centering
    \includegraphics[width=1.0\linewidth]{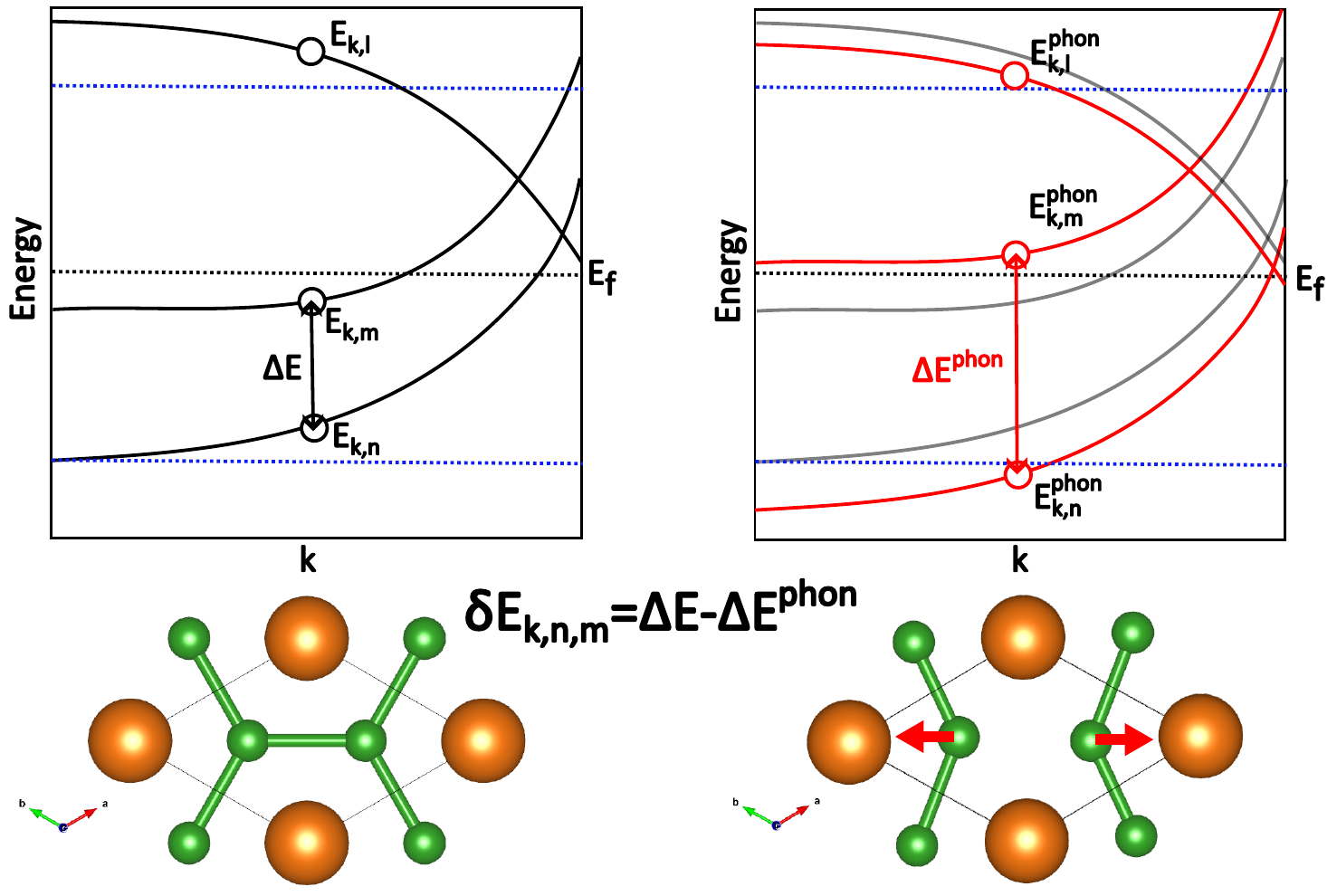}
\caption{\label{fig:deltaE_diagram} Explanatory diagram showing how $\partial E_{k,n,m,q,\nu}$ is calculated form the difference in electron energy band splittings for a phonon described by wave vector $q$ and band $\nu$. The blue dashed line indicates the energy window corresponding to the maximum phonon energy of that system.} 
\end{figure*}

Similarly to previous papers \cite{Yin2013,Hu2022} a frozen phonon approach can be taken to approximate
$\left \langle k+q \left| \delta V/ \delta u_{qj}^{\nu} \right|k \right \rangle$,
which is calculated from the size of the energy splitting of the bands at the Fermi surface. In fact it has been reported that this approach is able to account for non-adiabatic effects \cite{Hu2022} that can contribute up to 40$\%$ to the calculated values of $\lambda$. We approximate the electron-phonon matrix element (equation \ref{eq:M}) for a specific phonon mode of momenta $q$ and mode $\nu$ by the expression:
\begin{equation}
g_{k,n,m,q,\nu}= \frac{\partial E_{k,n,m,q,\nu}}{\partial x_{q,\nu}}\sqrt{\frac{\hbar}{2 M_0\omega_{q\nu}}} ,
\end{equation}
where $\partial x_{q,\nu}$ is the collective mass weighted phonon displacement length, $M_0$ is the total mass of the atoms in the cell, and $\omega_{q\nu}$ is the $q$-mode phonon frequency of band $\nu$. The phonon frequency is calculated using the frozen phonon method \cite{Lam1982} for each $q$-vector and mode $\nu$. In this method $\partial E_{k,n,m,q}$ is the change in the difference between the energies of electronic bands $m$ and $n$ at momenta $k$ between the equilibrium and frozen phonon system (i.e. what is the change in the energy splitting of the 2 bands, see Figure \ref{fig:deltaE_diagram}). It is evaluated at each $k$-point, and for bands $m$ and $n$, where the equilibrium system's electronic states' energies fall within an energy cutoff either side of the Fermi energy (chosen as the maximum phonon frequency, typically on the order of 100 meV). For single bands where $m=n$,  $\partial E_{k,n,n,q}$ is instead defined as the shift in the band energy with respect to the Fermi energy between the equilibrium and frozen phonon system. Given the small energy width of the cutoff only a small number of bands typically appear within the energy window. The greatest calculated $g_{k,n,m,q,\nu}$ = $g^{\textrm{max}}_{k,q,\nu}$ within the window is taken as the approximate value for the electron-phonon matrix element for that $k$, which can then be inserted into the full expression for $\lambda$ in order to compute $\lambda_{q,\nu}$ (the electron-phonon coupling for that specific phonon mode $\nu$ with momenta $q$).

Due to band folding in the supercells required to describe phonons with a given $q$, only vertical transitions between bands need to be calculated as $k+q$ folds onto $k$. 

The mode dependant electron-phonon coupling for a qiven $q$ is then approximated as
\begin{equation} \label{eq:lamda_poorman}
\lambda_{q,\nu} = \frac{2}{N\left(\epsilon_F\right)} \sum_{k,n,m}{\left|g^{\textrm{max}}_{k,q,\nu}\right|^2G\left(E_{k,n}-E_F\right)G\left(E_{k+q,m}-E_F\right)}
\end{equation}
where $n$ and $m$ are electronic band indices. As this is a sum over finite $k$- and $q$- grids, instead of using $\delta$-functions, Gaussians centred at the Fermi energy,
\begin{equation} \label{eq:gaussian}
G(E_k-E_F) = \frac{1}{\sigma\sqrt{2\pi}} \exp{\left(\frac{-(E_k-E_F)^2}{2\sigma^2}\right)}
\end{equation}
are used to smear the electronic band energies as is commonly done \cite{Shipley2020, Wierzbowska2005}. The Gaussian width, $\sigma$, is a tunable parameter that has a significant effect on the value of $\lambda$ and is present in codes like EPW\cite{Noffsinger2010}. If $\sigma$ is too small then the function becomes narrower than the average electron energy spacing in the $k$-grid and so will over emphasise some contributions whilst missing others, too broad and behaviour becomes too smeared and a poorer approximation the $\delta$-function.

To then calculate $\lambda_{q}$ the contribution from each mode is summed so that:
\begin{equation}
\lambda_q=\sum_{\nu}\lambda_{q,\nu}
\end{equation}

And then we calculate
\begin{equation}\label{eq:sumlambdaq}
\lambda=\frac{1}{N_q}\sum_{q}\lambda_q .
\end{equation}

Normally a uniformly dense grid of $q$-points would be used (with symmetries taken into consideration for calculation efficiency) in codes like EPW \cite{Noffsinger2010}, but here only selected high symmetry points are sampled. The main computational costs to this approach are the DFT calculations of the supercells that correspond to the particular phonon modes, with the poor man’s code itself executing on a single core in seconds or minutes. Significant speed up is achieved by not having to calculate the $\langle k+q| \delta V / \delta u_{qj}^{\nu} |k\rangle$ overlap integrals, and by limiting the number of $q$-points required.

\subsection{Fermi surface nesting}

The nesting function quantifies the amount to which certain $q$-vectors within the BZ connect points on a material's Fermi surface, and although its efficacy for predicting where charge density waves will occur is disputed \cite{Johannes2008}, the nesting function can tell us something about which $q$-vectors can (but not necessarily do) contribute to electron-phonon coupling in our model. Here we define it as

\begin{equation} \label{eq:nest_func}
\chi_{q}(E_F) = \frac{2}{N\left(\epsilon_F\right)} \sum_{k,n,m}{G\left(E_{k,n}-E_F\right)G\left(E_{k+q,m}-E_F\right)}
\end{equation}

using Gaussians instead of $\delta$-functions due to the finite $k$-mesh.

\subsection{Selecting a $q$-mesh}

Most of the systems compared to in this paper are smaller systems where larger $q$-meshes are possible using full Eliashberg methods, the benefits of the ``poor-man's" method is that it can be extended to extremely large systems. However, in order to compare and validate the method outlined above to other calculations, as large a $q$-mesh as possible has been used whilst still allowing reasonable calculation time for each system. To this end the estimated $\lambda$ of all materials has been calculated using all high symmetry phonon modes representable within a 2$\times$2$\times$2 supercell. The multiplicity of each mode has also been included when evaluating equation \ref{eq:sumlambdaq} so it becomes
\begin{equation}\label{eq:sumlambdaqmult}
\lambda=\frac{1}{M_q N_q}\sum_{q} M_q \lambda_q ,
\end{equation}
where $M_q$ is the multiplicity of the $q$-point in the BZ.

\section{\label{sec:Results} Results and Discussion}

\subsection{Applying the ``poor-man's" approximation to MgB$_{2}$}

MgB$_2$ is an ideal test system as it is a BCS phonon-mediated superconductor with a high $T_{c}$ ~39 K \cite{Nagamatsu2001} and the electron-phonon coupling has been calculated previously using the Eliashberg equations from DFT calculated wavefunctions, with values of 0.643 \cite{Johansson2022}, 0.73 \cite{Bohnen2001} and independently 0.73 \cite{Morshedloo2015}. From these calculations it is well known which phonon modes contribute significantly to its $\lambda$, specifically at the $\Gamma$-point. This means not only can we compare the poor-man's predicted values of lambda, but we can determine whether the method is able to pick up on the correct response of the electronic bands to specific phonon modes.

\begin{figure}
\centering
    \includegraphics[width=1.0\linewidth]{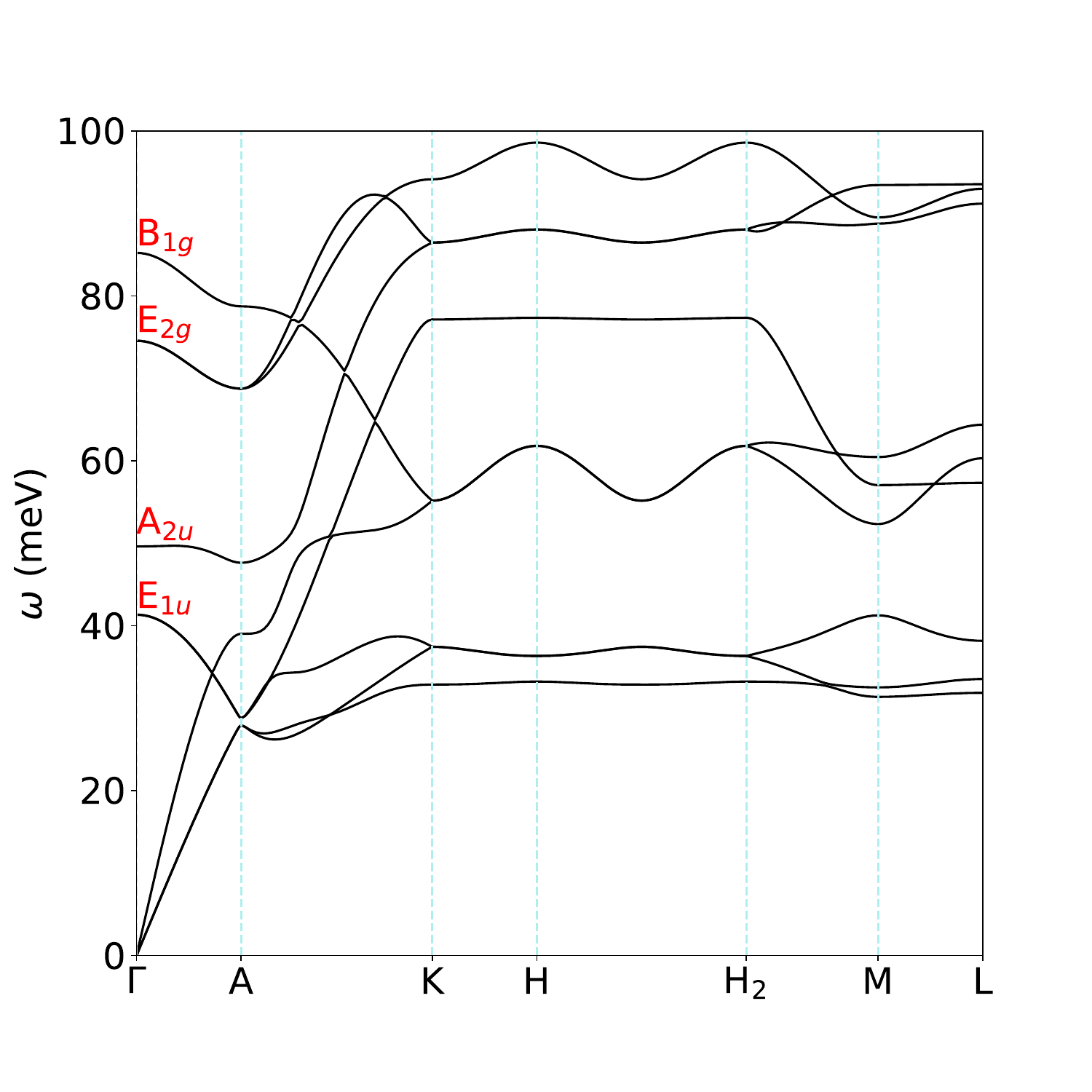}
\caption{\label{fig:MgB2_phonbands} Phonon band diagram of MgB$_2$ with the $\Gamma$-modes labelled.} 
\end{figure}

In a recent paper by Sun et al. \cite{Sun2022} a method is used to estimate the electron-phonon coupling in MgB$_2$, calculating only $\Gamma$-centered modes and assuming a uniform $g$ factor over the Brillouin zone (BZ), and then including a fudge factor to account for differences in nesting over the BZ. As can be seen in Figure \ref{fig:MgB2_phonbands}, of the 9 phonon modes, there are only 4 phonon modes distinct in energy at $\Gamma$, with some modes being degenerate. It is well known from frozen phonon \cite{Sun2022} and Eliashberg \cite{Bohnen2001,Johansson2022,Morshedloo2015} calculations that the phonon mode that contributes most strongly to EPC in MgB$_2$ is the $\Gamma$-point E$_{2g}$ mode which perturbs the B-B $\sigma$-bond \cite{Mazin2002}. Sun et al. \cite{Sun2022} indeed show that out of those 4 modes only the E$_{2g}$ mode contributes to the electron-phonon coupling, with no other modes showing perturbation to the unscreened phonon frequencies.

\begin{table}[]
\begin{tabular}{cllll}
\hline
\thead{Gaussian \\ Width (meV) }& $\lambda_{E_{1u}}$  & $\lambda_{A_{2u}}$& $\lambda_{E_{2g}} $ & $\lambda_{B_{1g}}$  \\
\hline
\hline
        13.6 & 0.000 & 0.001 & 0.652 & 0.003 \\ 
        27.2 & 0.000 & 0.000 & 0.597 & 0.002 \\ 
        40.8 & 0.000 & 0.000 & 0.553 & 0.001 \\ 
        54.4 & 0.000 & 0.000 & 0.467 & 0.001 \\ 
        68.0 & 0.000 & 0.000 & 0.379 & 0.001 \\ 
        81.6 & 0.000 & 0.000 & 0.306 & 0.001 \\ 
        95.2 & 0.000 & 0.000 & 0.248 & 0.000 \\ 
        108.8 & 0.000 & 0.000 & 0.203 & 0.000 \\ 
        122.5 & 0.000 & 0.000 & 0.168 & 0.000 \\ 
\hline
\end{tabular}
\caption{\label{table:MgB2_lamdaq} The contribution to $\lambda_{\Gamma}$ (the EPC of the $\Gamma$ phonon mode) of each distinct $\Gamma$ phonon mode compared to the Gaussian width (see equation \ref{eq:lamda_poorman}). The only substantial contribution comes from the E$_{2g}$ mode which is doubly degenerate.} 
\end{table}

\begin{figure}
\centering
    \includegraphics[width=1.0\linewidth]{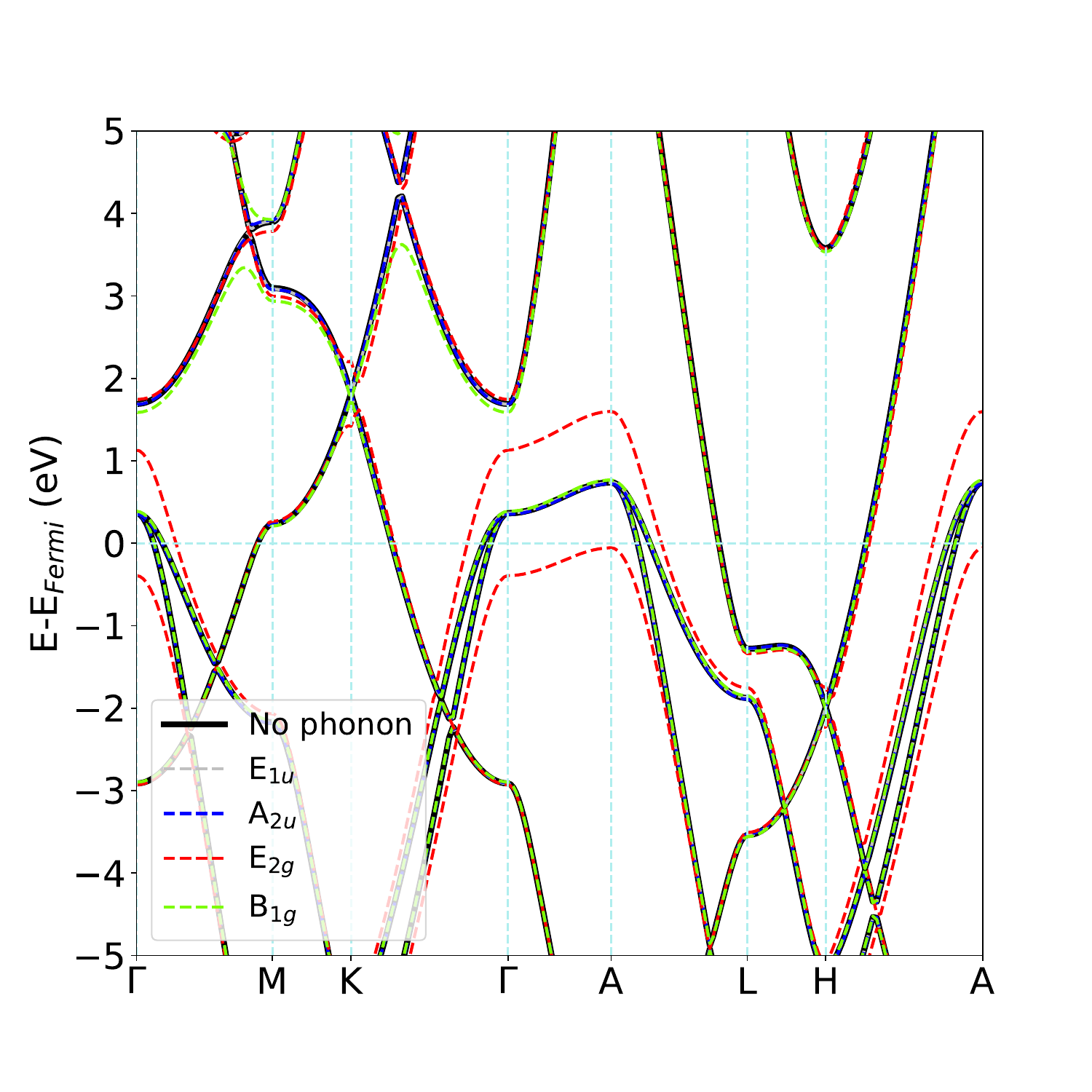}
\caption{\label{fig:MgB2_ebands} Electronic band diagrams of the $\Gamma$ frozen phonons in MgB$_2$ compared to the band structure of the unperturbed system. It becomes clear that only the E$_{2g}$ phonon strongly perturbs the electronic bands near the Fermi energy and so will be the only band that contributes to the EPC.} 
\end{figure}

As can be seen in Table \ref{table:MgB2_lamdaq}, our poor-man’s calculations of $\lambda$ agrees with these previous results \cite{Bohnen2001,Sun2022,Johansson2022,Morshedloo2015}, in that only the E$_{2g}$ mode at the $\Gamma$ point significantly contributes to $\lambda$, for all values of Gaussian width (our poor man’s $\lambda$ was calculated using an energy window equal to the maximum phonon energy, 99 meV, either side of the Fermi energy, with the DFT calculations run with a 24×24×24 $k$-grid). The reason for the strong response due to this mode can be seen in the electronic band structure (see Figure \ref{fig:MgB2_ebands}), where the E$_{2g}$ phonon mode is the only one that perturbs the electronic band structure of MgB$_2$ near the Fermi energy. A strong splitting in the B-B $\sigma$ bonding band between $\Gamma$ and A where the band crosses the Fermi energy can be observed for displacements of the 2 B ions of only 0.062 \AA. A similar effect of the E$_{2g}$ frozen phonon on this specific band has been demonstrated previously \cite{Zhai2019}, but not used to predict a value for the EPC. This behaviour demonstrates that our proposed method to use the perturbations of electronic band states with energies very close to the Fermi energy (of the order of the phonon energies) due to frozen phonons can be used to predict where there will be strong electron-phonon coupling.

The issue then becomes whether this qualitative result can be used to make a quantitative prediction of $\lambda$. The predicted value of $\lambda$ is dependent on the chosen Gaussian width, but for a width of 0.06 Ry (81.6 meV) (which in the below section gives us the best overall fit to all the data) the predicted $\lambda$ is calculated to be 0.65 after averaging over 4 high symmetry $q$-points. In this case $\lambda_{q=\Gamma}$ is approximately equal to  $\lambda$ over the whole BZ and would suggest only the $\Gamma$-mode need be calculated, however this is found to not generally be the case.

\subsection{General predictions of $\lambda$}

\begin{figure}
\centering
    \includegraphics[width=1.0\linewidth]{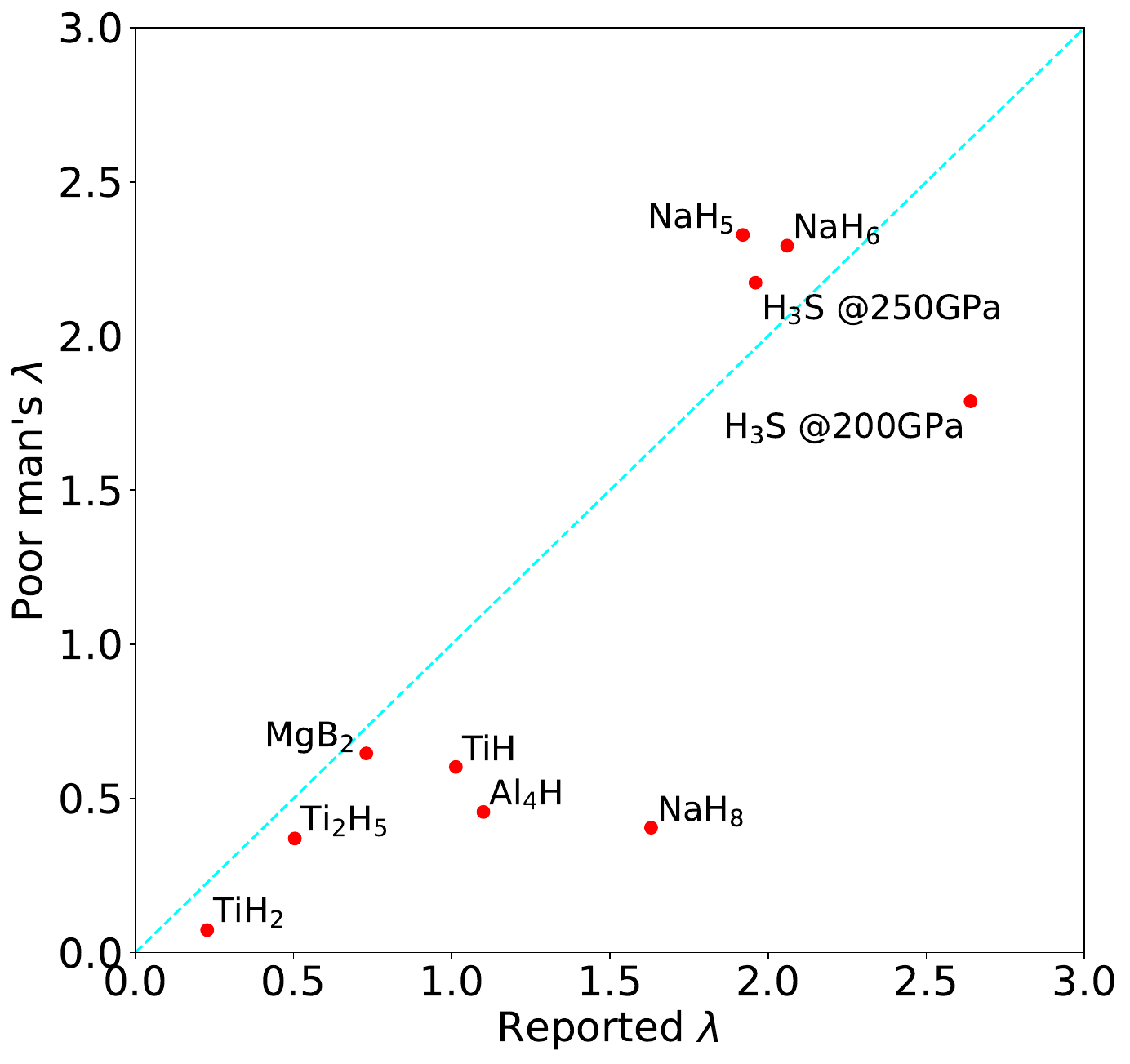}
\caption{\label{fig:Lambda_scatter} A scatter plot of $\lambda$ calculated using the approximation described in this paper and a Gaussian width of 0.06 Ry (81.6 meV) against that reported in the literature \cite{Errea2015,Bohnen2001,Zhang2020,Shipley2021,He2023}.} 
\end{figure}

In Fig. \ref{fig:Lambda_scatter} a summary of the $\lambda$ values can be seen plotted against reference values estimated by our method for MgB$_2$ \cite{Bohnen2001}, H$_3$S \cite{Errea2015}, series of titanium hydrides \cite{Zhang2020} and sodium hydrides \cite{Shipley2021}, and a newly predicted Al$_4$H structure \cite{He2023}. The predicted values were calculated using a Gaussian width of 0.06 Ry (81.6 meV) which was the value that provided the best linear fit.

Comparing to a variety of literature values is complicated by the fact that they use different implementations to calculate EPC, including the use of different pseudopotentials, $k$-mesh densities, or smearing values for the approximations of the $\delta$-functions. However, references have been selected that are relatively consistent. All use the PBE functional, similar k-mesh densities for electronic structure calculations and, with the exception of H$_3$S \cite{Errea2015} and MgB$_2$ \cite{Bohnen2001}, all use Quantum ESPRESSO DFPT or EPW to calculate the EPC. However, this still allows for some variation between references even if were they to calculate the same materials, and thus also with the ``poor-man's" approach.

Even with these variations, when we plot our estimates for $\lambda$ against the literature values in Fig. \ref{fig:Lambda_scatter} the fit shown  is sufficiently good to at least provide a pre-screening estimate for $\lambda$. The coefficient of determination (R$^2$) value is 0.43 for a linear fit of predicted to reference values, when a Gaussian width of 0.06 Ry (81.6 meV) is used. The greatest outlier is the electron-phonon coupling constant predicted for NaH$_8$, which significantly reduces the goodness of fit. If this data point is removed the R$^2$ value rises to 0.73 for a linear fit. This is comparable to previously reported frozen-phonon methods, with Sun et al. \cite{Sun2022} achieving R$^2$=0.55 only when a ``fudge-factor" of 0.22 is applied, meaning the contributions they calculated from $\lambda_{q=\Gamma}$ underestimate the reported $\lambda$ by a factor of ~5. This demonstrates the limitation with approximating a uniform contribution to $\lambda$ across the whole BZ when nesting in many superconducting materials is strongly peaked.

\begin{figure}
\centering
    \includegraphics[width=1.0\linewidth]{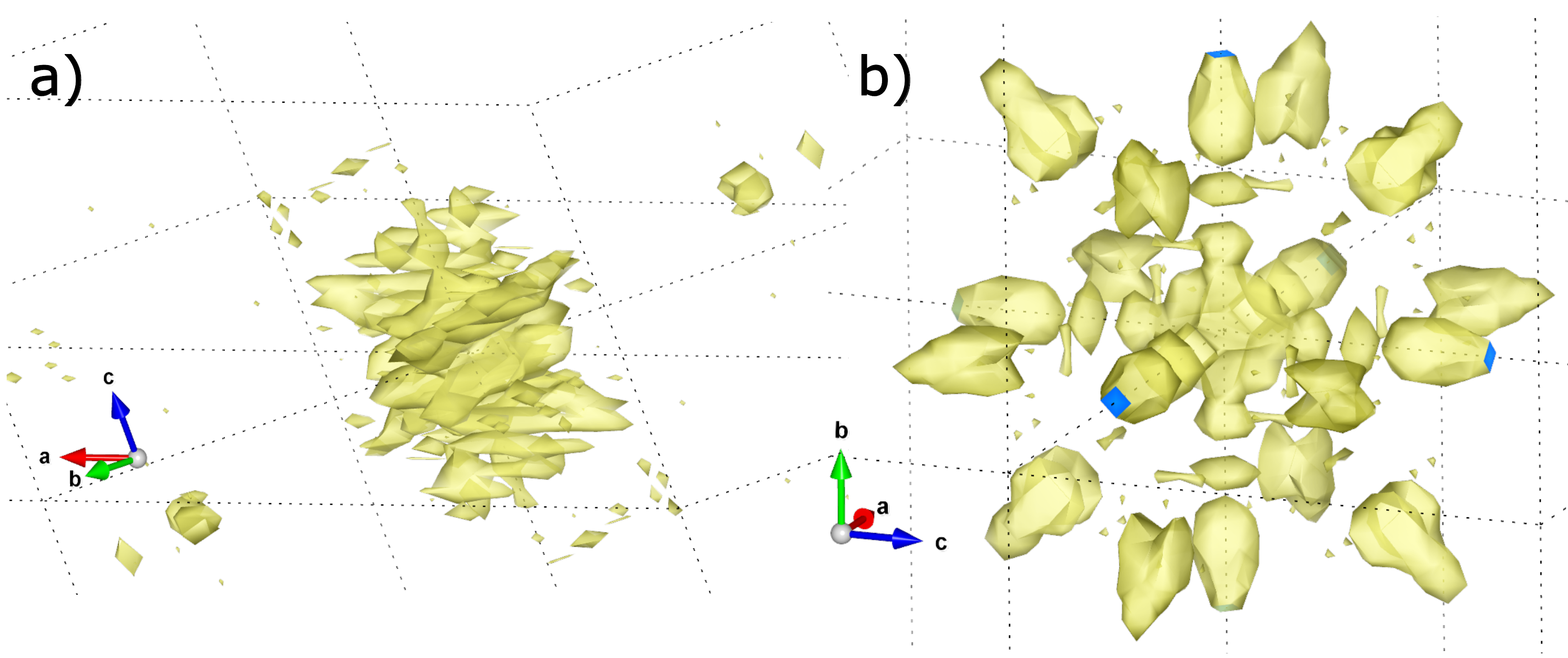}
\caption{\label{fig:Na_hydride_nest} Isosurfaces of the nesting functions of a) NaH$_8$ and b) NaH$_6$} 
\end{figure}

The under-estimation of $\lambda$ calculated for NaH$_8$ is likely due to the selected $q$-points of the frozen phonons that make up the $q$-mesh  not adequately sampling the distribution of $\lambda{_q}$ over the full BZ, with the contributions from phonon modes lying on the high symmetry points available in a 2$\times$2$\times$2 supercell over-representing areas with low contributions to $\lambda$. In the reference data \cite{Shipley2021} a denser $q$-mesh is used and, as can be seen in Figure \ref{fig:Na_hydride_nest}, the nesting function calculated for NaH$_8$ shows that the strongly nested regions of the BZ are concentrated around $\Gamma$ and along the $q_{z}$ axis in contrast. This is in contrast to NaH$_6$ where the areas of strong nesting are more evenly distributed across the BZ and at the high symmetry points at the BZ boundary, and we correspondingly estimate a more accurate value for $\lambda$.

\subsection{H$_{3}$S}

H$_3$S at high pressure (200 GPa) is a high temperature superconductor with a $T_{c}$ experimentally measured to be 203 K \cite{Drozdov2014,Drozdov2015}. It has an exceptionally high electron-phonon coupling strength calculated to be 2.64 \cite{Errea2015} in the harmonic approximation and has been well studied using ab initio techniques \cite{Ge2016}. These extensive studies of EPC in H$_3$S mean that we can compare our ``poor-man's" approximation in detail to more computationally expensive approaches.

\begin{figure}
\centering
    \includegraphics[width=1.0\linewidth]{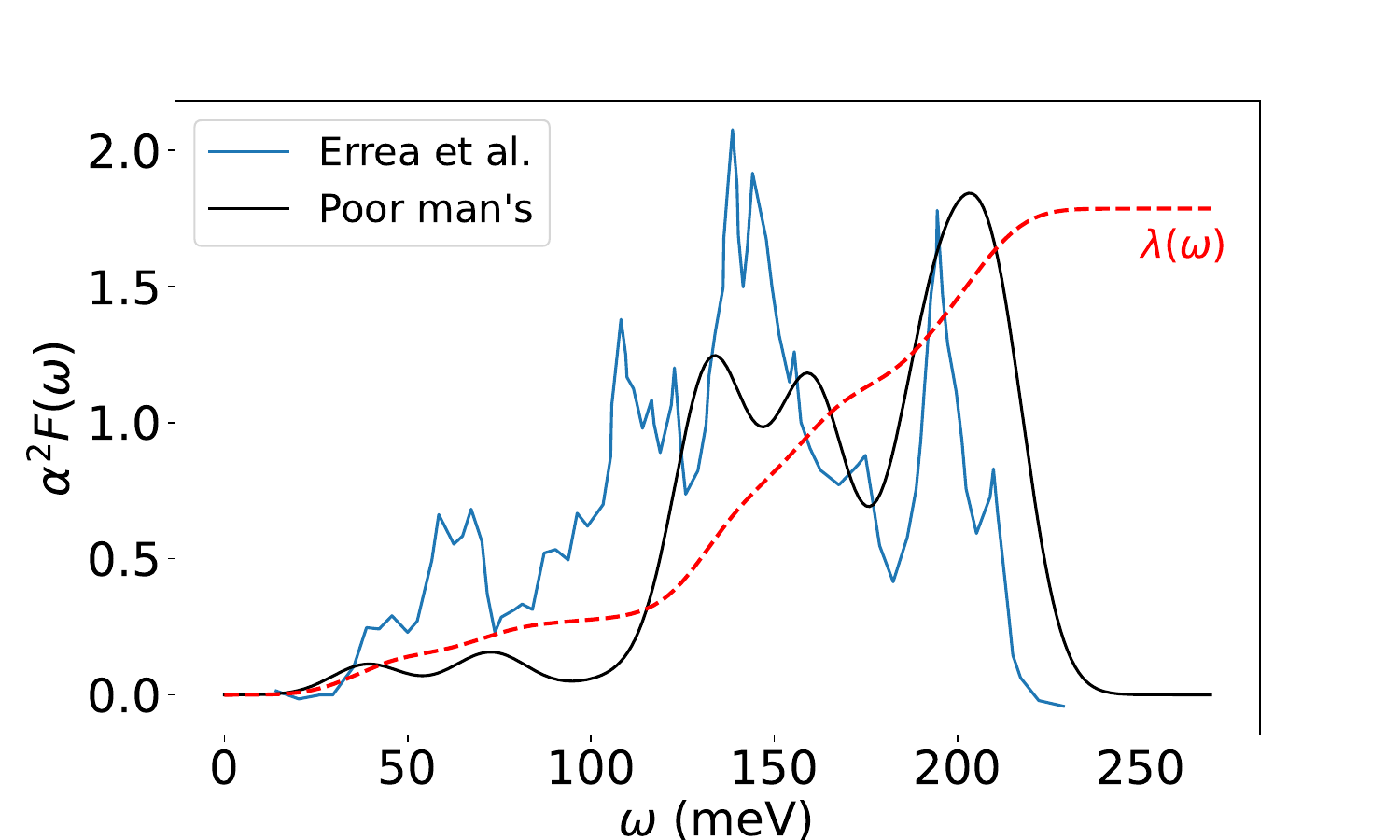}
\caption{\label{fig:H3S_alpha2F} The $\alpha^2F(\omega)$ function approximated from the poor man's approach for H$_3$S at 200 GPa using a 2$\times$2$\times$2 $q$-mesh compared to that reported by Errea et al. \cite{Errea2015} where a 6$\times$6$\times$6 $q$-mesh was used. The red dashed line shows $\lambda(\omega)$ calculated from $\alpha^2F(\omega)$. } 
\end{figure}

Using our poor man’s code we have calculated $\lambda_{q,\nu}$ for the 3 high symmetry $q$-points at (0,0,0), ($\frac{1}{2}$,0,$\frac{1}{2}$) and ($\frac{1}{2}$,$\frac{1}{2}$,$\frac{1}{2}$). From these $\lambda_{q,\nu}$ and their corresponding phonon frequencies, $\omega_{q\nu}$, we can also obtain  an estimate for the isotropic Eliashberg function 

\begin{equation}
\alpha^2F(\omega)=\frac{1}{2N_{q}}\sum_{q\nu}\lambda_{q\nu}\omega_{q\nu}\delta(\omega-\omega_{q\nu})
\end{equation}

which we can compare against previous results \cite{Errea2015} (see Figure \ref{fig:H3S_alpha2F}). Here we have approximated $\delta$-function using a Gaussian function with a width of 20 meV. We also plot $\lambda(\omega)=2\int^{\omega}_{0}\alpha^2F(\omega')/\omega' d\omega'$ which gives $\lambda(\omega)=\lambda$ (the isotropic EPC) as $\omega \to \infty$. As can be seen in Figure \ref{fig:H3S_alpha2F} our approximation of $\alpha^2F(\omega)$ is a good approximation of that calculated using a denser 6$\times$6$\times$6 $q$-mesh and fully calculating the electron-phonon matrix elements. Our underestimation of $\lambda$ can be explained when comparing to previous papers\cite{Errea2015,Ge2016} which calculate a strong contribution to $\lambda$ for phonon modes with $q$ halfway between $\Gamma$ and P, ($\frac{1}{4},\frac{1}{4},\frac{1}{4}$). This would correspond to an 8$\times$8$\times$8 supercell with 2,048 atoms which becomes unfeasibly large to compute using a frozen phonon approach. However, for the purposes of screening a much reduced $q$-mesh is sufficient.

\section{\label{sec:Conclusion} Conclusion}

We have proposed a computationally inexpensive frozen phonon method to estimate the electron-phonon coupling, $\lambda$, of materials whilst taking into account the effects of nesting and the electronic band structure. This approach yields sufficient accuracy that it can be used for screening promising materials in high-throughput searches for high temperature superconductors.

Our approach also opens up the possibility of screening large primitive cell structures or 2-dimensional films and interfaces for enhanced superconductivity. With more data it might be possible to predict or learn a priori which phonon modes (or atomic displacements) are most likely to contribute to strong coupling thus cutting computational cost further and allowing testing of ever larger and more complex systems.

\section{Acknowledgments}

We are grateful to Philipp Brauninger-Weimer and Nassim Derriche for useful discussions. This work was supported by the Deep Science Fund at Intellectual Ventures and the Engineering and Physical Sciences Research Council (EP/X034429/1). This research was undertaken thanks, in part, to funding from the Max Planck-UBC-UTokyo Center for Quantum Materials and the Canada First Research Excellence Fund, Quantum Materials and Future Technologies Program as well as from CIFAR and NSERC.

\bibliography{Electron-phonon-coupling}

\end{document}